\documentclass[twocolumn,prl,aps,floats]{revtex4}

\usepackage{amssymb,epsf,graphicx,xcolor}
\usepackage{supertabular}
\usepackage{epsfig,amsmath}
\usepackage{xfrac}
\usepackage{amsfonts}
\usepackage{mathtools}
\usepackage{cleveref}

\newcommand{\ov}{\overline}

\begin{document}

\frenchspacing
\title{Number of attractors in the critical Kauffman model is exponential}
\author{T. M. A. Fink and F. C. Sheldon}
\affiliation{London Institute for Mathematical Sciences, Royal Institution, 21 Albemarle St, London W1S 4BS, UK}

\date{\today}
\begin{abstract}
\noindent
The Kauffman model is the archetypal model of genetic computation.
It highlights the importance of criticality, at which many biological systems seem poised.
In a series of advances, researchers have honed in on how the number of attractors in the critical regime grows with network size.
But a definitive answer has proved elusive.
We prove that, for the critical Kauffman model with connectivity one, the number of attractors grows at least, and at most, as $(2/\!\sqrt{e})^N$.
This is the first proof that the number of attractors in a critical Kauffman model grows exponentially.
\end{abstract}

\maketitle
\noindent
\noindent
{\sf\textbf{\textcolor{purple}{Introduction}}} \\
The Kauffman model is a discrete dynamical system on a random directed graph of $N$ nodes, each with $K$ inputs but any number of outputs~\cite{Kauffman69a,Kauffman69b}. 
The state of each node is a Boolean function of the states of its neighbors, 
and all nodes are updated simultaneously.
Despite its simplicity, the system captures essential features of many physical systems, such as gene regulation~\cite{Wang12,Ahnert16}, chemical reaction networks~\cite{Hordijk04}, and economic networks~\cite{Easton08}. 
It is particularly useful as a null model, whereby its baseline behavior can be contrasted against networks that have specific structure.
\\ \indent
The long-term behavior of the Kauffman network falls into two regimes~\cite{Aldana03,Drossel08}. 
In the frozen regime, perturbations to the initial state die out and attractor lengths do not grow with system size.
In the chaotic regime, perturbations can grow exponentially and attractor lengths grow with the size of the system size.
The two regimes are separated by a critical boundary in which a perturbation to one node spreads to on average one other node \cite{Derrida86a,Derrida86b,Aldana03}.
The properties of this critical region are interesting to scientists across a broad range of fields~\cite{Munoz18,Daniels18}.
\\ \indent
For the critical $K=2$ model, early computational evidence \cite{Kauffman69b}
indicated that the mean number of attractors scaled as $\sqrt{N}$, where $N$ is the size of the network. 
But complete enumeration up to $N=32$ suggested linear scaling~\cite{Bilke01}.
Socolar and Kauffman found evidence for faster than linear~\cite{Socolar03}, noting that the correct scaling only emerges for large systems.
Through a combination of simulation and arguments about modular structure, Bastola and Parisi suggested a stretched exponential \cite{Bastola98a,Bastola98b}.
Then, in an analytical tour de force, Samuelsson and Troein~\cite{Samuelsson03} proved that the number of attractors grows faster than any power law.
\begin{figure}[b!]
\noindent
      \epsfxsize=1 \columnwidth
      \epsfbox{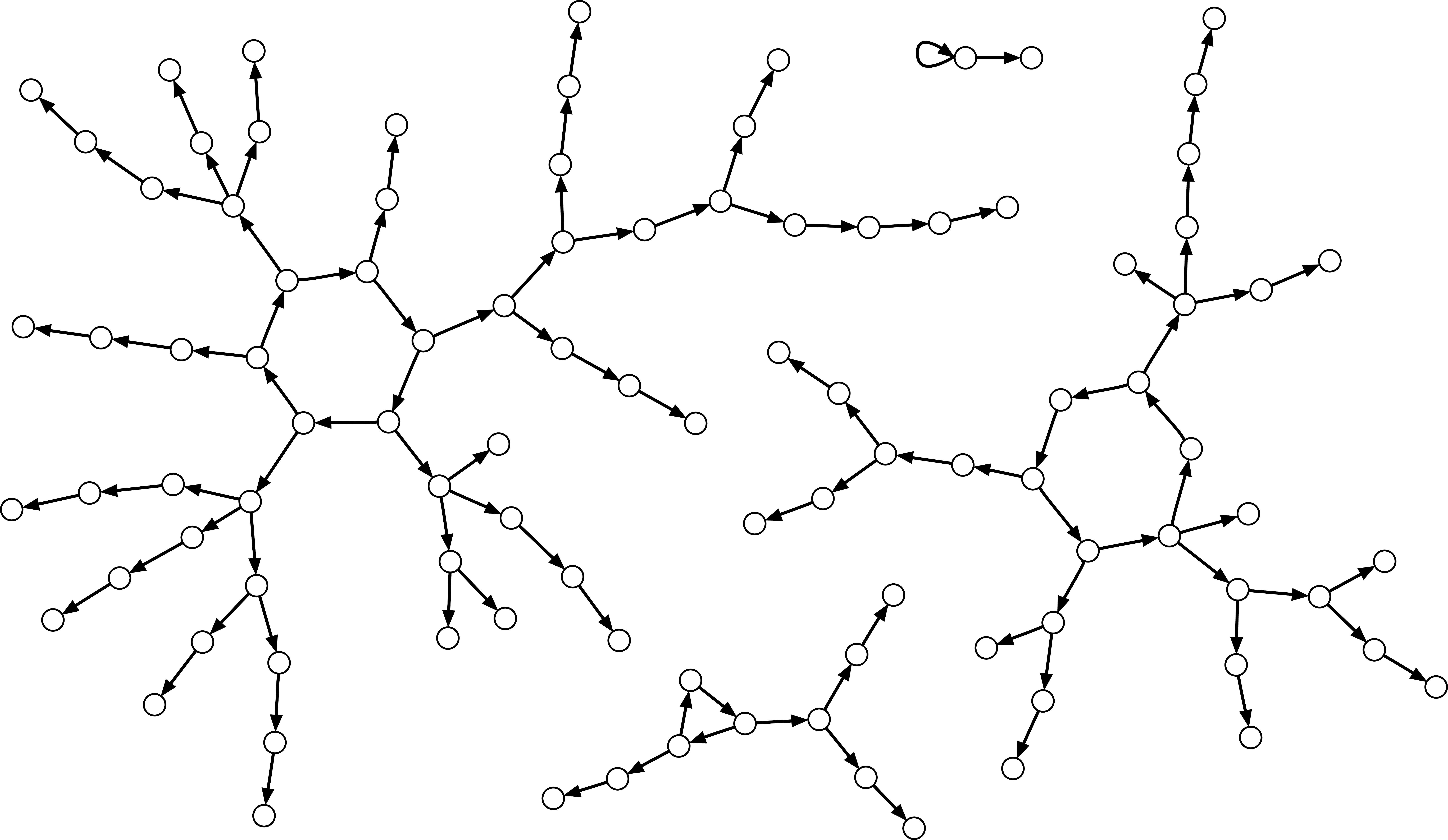}
\begin{small}
\caption{
{\bf Kauffman network with connectivity \boldmath$K=1$.} 
This typical network of $N=100$ nodes has one 1-loop, one 3-loop and two 6-loops, for a total of $m=16$ nodes in loops.
}
\end{small}
\end{figure}
\\ \indent
The elusive scaling for the $K=2$ motivated interest in the scaling for $K=1$.
This model has a beautifully simple structure that makes it amenable to analytical work.
The network is composed of loops and trees branching off the loops, shown in Fig. 1. 
Because the nodes in the trees are determined by the nodes in loops, they do not contribute to the number or length of attractors, which are set solely by the $m$ nodes in the loops. 
Every node is assigned one of four Boolean functions: on, off, copy or invert.
However, the critical version of the problem requires that all Boolean functions in the loops be copy or invert \cite{Flyvbjerg88,Drossel05a,Fink23a,Fink23b}.
\\ \indent
For the critical $K=1$ model,
Flyvbjerg and Kjaer found that the number of attractors is at least $2^{0.63 \sqrt{N}}$ \cite{Flyvbjerg88}.
Drossel \emph{et al.} obtained a slightly slower growth rate of $2^{0.59 \sqrt{N}}$, but with a much simpler calculation \cite{Drossel05a}.
We recently improved this to $2^{1.25 \sqrt{N}}$, by defining a product between dynamics in different loops \cite{Fink23a}. 
We obtained a similar result with a simpler calculation using results from number theory \cite{Fink23b}.
\\ \indent
In this paper, we prove that the number of attractors in the critical Kauffman model with connectivity one 
grows at least, and at most, as $\left( 2/\!\sqrt{e} \right)^N$,
where $N$ is the network size and $2/\!\sqrt{e} = 1.213$.
This is the first proof that the number of attractors in a critical Kauffman network grows exponentially.
Our approach is as follows.
We start by working out the exact probability that $m$ of the $N$ network nodes are in loops.
For a given $m$, we write down the minimum and maximum number of attractors that the network can have.
We then average these quantities over $m$.
To our surprise, the two averages converge to $\left( 2/\!\sqrt{e} \right)^N\!$.
Our proof is in four steps, each of which relies on a lemma. 
We prove the four lemmas after the derivation of our main result.
\\ \\ \noindent
{\sf\textbf{\textcolor{purple}{Proof that number of attractors is exponential}}}\\
We now prove that the mean number of attractors scales as $(2/\!\sqrt{e})^N$.
\\ \indent
{\emph{\textcolor{purple}{Step 1.}}} 
The probability that $m \in [1,N]$ of the network nodes are in loops is
\begin{equation*}
P(m) = \frac{m}{N} \frac{N!}{(N-m)!} \frac{1}{N^m},
\end{equation*}
which we prove in lemma 1 below.
\\ \indent
For a given $m$, the maximum number of attractors occurs when all of the loops are even and of length 1.
Then there are $2^m$ attractors, all of size 1.
On the other hand, the largest attractor length is double the lcm of the individual loop sizes.
This is precisely twice Landau's function.
From our previous work \cite{Fink23b}, we know that the minimum number of attractors is $2^{m - 1.52 \sqrt{m \ln m}}/2$,
in which we made use of a bound on Landau's function \cite{Massias85}.
Thus
\begin{equation*}
    c(m) 
    \begin{dcases}
        \geq    
        2^{m - 1.52 \sqrt{m \ln m}}/2, \\
        \leq       
        2^m.
    \end{dcases}
    \label{BoundsA}
\end{equation*}
By summing this over the distribution of $m$, we can write down bounds for the mean number of attractors $\overline{c}(N)$:
\begin{equation}
    \overline{c}(N) 
    \begin{dcases}
        \geq    
        \frac{1}{2} \sum_{m=1}^N 2^{m - 1.52 \sqrt{m \ln m}} P(m), \\
        \leq       
        \sum_{m=1}^N 2^m P(m).
    \end{dcases}
    \label{BoundsB}
\end{equation}
\indent
{\emph{\textcolor{purple}{Step 2.}}} 
The quantity $\sqrt{m \ln m}$ is difficult to handle analytically, but we can bound it from above by a line in $m$. 
Let $\epsilon \in (0,1]$ be an arbitrarily small constant.
As we prove in lemma 2 below,
\begin{equation*}
m \epsilon + \frac{b^2}{\epsilon} \ln\left(\frac{b}{\epsilon}\right) > b \sqrt{m \ln m},
\end{equation*}
where $b = 1.52$.
Inserting this into eq. (\ref{BoundsB}) gives
\begin{equation}
    \overline{c}(N) 
    \begin{dcases}
        > \overline{c}_{\rm min}(N) =    
        \frac{2^{b^2\!/\epsilon \, \ln(\epsilon/b)}}{2} \sum_{m=1}^N 2^{m(1 - \epsilon)} P(m), \\
        \leq       
        \overline{c}_{\rm max}(N) = 
        \sum_{m=1}^N 2^m P(m).
    \end{dcases}
    \label{BoundsC}
\end{equation}
\\ \indent
{\emph{\textcolor{purple}{Step 3.}}} 
It turns out that these sums can be calculated exactly for finite $N$.
We show how in lemma 3 below.
In the limit of large $N$, the exact form becomes
\begin{equation}
    \overline{c}(N)
    \begin{dcases}
        >    
        \frac{\sqrt{2 \pi N}}{2}
        \left(1 \!-\! \textstyle{\frac{2^\epsilon}{2}}\right)
        \left(\frac{\textstyle{\frac{2}{2^\epsilon}}}{\exp(1\! -\! \textstyle{\frac{2^\epsilon}{2}})} \right)^{\!N} \!\!\!
        \left(\frac{\epsilon}{b}\right)^{b^2/\epsilon \ln 2}, \\
        <       
        \frac{\sqrt{2 \pi N}}{2} 
        \left(\frac{2}{\sqrt{e}}\right)^N\!.
    \end{dcases}
    \label{BoundsD}
\end{equation}
\begin{figure}[b!]
\noindent
      \epsfxsize=1 \columnwidth
      \epsfbox{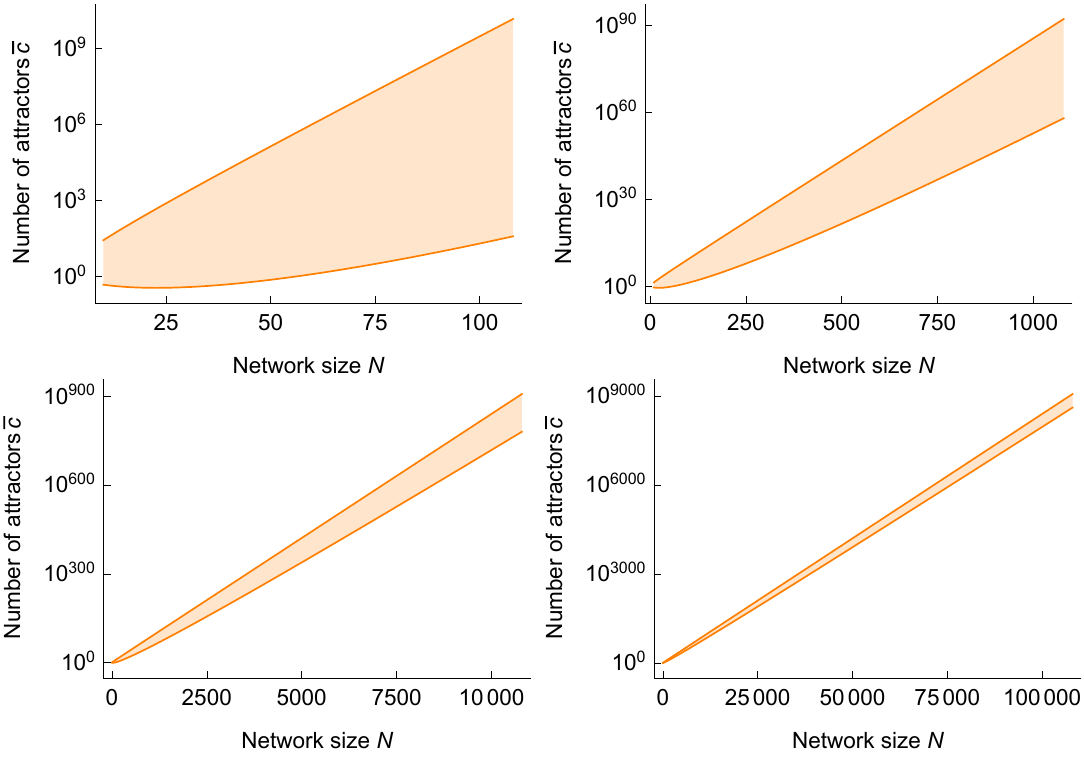}
\begin{small}
\caption{
{\bf The mean number of attractors \boldmath${\overline{c}(N)}$ grows as \boldmath${(2/\!\sqrt{e})^N}$.} 
As the network size $N$ increases, the slope of the lower bound approaches the slope of the upper bound,
namely, $\ln (2/\!\sqrt{e})$.
}
\end{small}
\end{figure}
For small $\epsilon$, this simplifies to 
\begin{equation}
    \overline{c}(N)
    \begin{dcases}
        >    
        1.25 \sqrt{N}
        \left(\frac{2 - \epsilon \ln 2}{\sqrt{e}}\right)^{\!N} \!\!\!
        \left(\frac{\epsilon}{b}\right)^{b^2/\epsilon \ln 2}, \\
        <       
        1.25 \sqrt{N}
        \left(\frac{2}{\sqrt{e}}\right)^N\!.
    \end{dcases}
    \label{BoundsE}
\end{equation}
The large $N$ and small $\epsilon$ approximations are rather accurate.
For example, for $\epsilon = 0.01$ and $N = 10^5$, 
eq. (\ref{BoundsC}) gives for the log of the lower bound 18,170,
eq. (\ref{BoundsD}) gives 18,170, 
and eq. (\ref{BoundsE}) gives 18,169.
\\ \indent
We can make $(2 - \epsilon \ln 2)/\sqrt{e}$ as close as $2/\sqrt{e}$ as we wish.
Doing so makes the constant on the right side of the lower bound smaller.
The essential point is that this constant is independent of $N$.
We thus see that the number of attractors scales at least, and at most, as $\left( 2/\!\sqrt{e} \right)^N$.
\\ \indent
{\emph{\textcolor{purple}{Step 4.}}} 
The rate at which these bounds converge to the same scaling can be determined explicitly.
For any $N$, there is a choice of $\epsilon$ which maximizes the lower bound. 
Setting $\epsilon$ to this optimal value allows us to get rid of $\epsilon$ altogether.
Our result, which we derive in lemma 4, is
\begin{equation}
   \ln \overline{c}(N) 
    \begin{dcases}
        > \! N \ln \! \left(\frac{2}{\sqrt{e}}\right) 
        \left(\! 1-\frac{5.45}{\sqrt{N}} \frac{(W(e^2N)-1)}{\sqrt{W(e^2 N)}}
        \right)\!, \\
        <  \! N \ln \! \left(\frac{2}{\sqrt{e}}\right),
    \end{dcases}
    \label{BoundsF}
\end{equation}
where $W$ is the principal branch of the Lambert $W$ function.
This approach is plotted in Fig. 2.
Since $W(x)$ asymptotically grows as $\ln x$,
the right side of the lower bound approaches 1 as $1-\sqrt{\ln N / N}$.
For example, for $N = 10^6$ and $10^9$, the right side is 0.978 is 0.999.
Once again, we find that the number of attractors scales as $(2/\!\sqrt{e})^N$.
\\ \\ \noindent
{\sf\textbf{\textcolor{purple}{Proof of our four lemmas}}} \\
{\emph{\textcolor{purple}{Lemma 1:  Distribution for $m$.}}} 
We aim to directly count the number of network architectures which have $m$ nodes in loops. We consider the case of labelled nodes. 
Take $m$ of $N$ nodes $\binom{N}{m}$ ways and arrange them in cycles in $m!$ ways. For the remaining $N-m$ nodes not to create loops, they must be arranged in trees protruding from the loops.
\\ \indent
Given a set of $k$ trees, there are $m^k$ ways of attaching them to the $m$ loop nodes. Calling the number of labelled, rooted $k$-forests of $N-m$ nodes $F(N-m, k)$, we sum over $k \in 1...N\!-\!m$ to find the total number of arrangements of nodes in trees attached to the loop nodes as
\begin{equation*}
    \sum_{k=1}^{N-m} m^k F(N-m, k).
\end{equation*}
From Moon~\cite{Moon70}, $F(n, k) = \binom{n}{k} k n^{n-k-1}$ which allows us to evaluate the sum above as,
\begin{equation*}
    \sum_{k=1}^n m^k \binom{n}{k} k n^{n-k-1} = m(m+n)^{n-1}.
\end{equation*}
Putting the pieces together and dividing by the total number of networks $N^N$ we have
\begin{equation*}
    P(m) =\binom{N}{m}m!\frac{m N^{N-m-1}}{N^N}= \frac{m}{N}\frac{N!}{(N-m)!N^m}.
\end{equation*}
For example, for $N=3$, this gives $3/9, 4/9, 2/9$ for $m = 1,2,3$.
Note in particular that the last term is $N!/N^N$.
This makes sense, since there are $N^N$ ways of drawing a single-input network on $N$ nodes, but $N!$ of these are permutations, that is to say, have all $N$ nodes in loops.
\\ \indent
{\emph{\textcolor{purple}{Lemma 2: Bound on $\sqrt{m \ln m}$.}}} 
Now we prove that
\begin{equation}
m \epsilon + \frac{b^2}{\epsilon} \ln\left(\frac{b}{\epsilon}\right) > b \sqrt{m \ln m},
\label{KB}
\end{equation}
where $\epsilon \in (0,1]$ and $b$ is a parameter which we will later set to $1.52$.
We want to find a line $m \epsilon + w(\epsilon)$ such that
\begin{equation*}
    \epsilon m + w(\epsilon) > b \sqrt{m\ln m},
\end{equation*}
where $\epsilon$ is the slope and $w(\epsilon)$ is a constant that depends on $\epsilon$ but is independent of $m$.
Let's first transform this equation to combine $\epsilon$ and $b$ into one variable.
With $\delta = \epsilon/b$ and $v(\delta) = w(\epsilon)/b$, we have
\begin{equation*}
    \delta m + v(\delta) > \sqrt{m\ln m}.
\end{equation*}
Our ansatz is $v(\delta) = - \ln \delta/\delta$, in which case
\begin{equation}
    \delta m - \frac{\ln \delta }{\delta} > \sqrt{m\ln m}.
    \label{KM}
\end{equation}
Squaring both sides and rearranging,
\begin{equation*}
    \delta^2 m + \frac{\ln^2 \delta }{\delta^2 m} > \ln (\delta^2 m).
\end{equation*}
With $u = \delta^2 m$, we can write
\begin{equation*}
    e^u \exp\left(\frac{\ln^2 \delta}{u}\right) > u,
\end{equation*}
where $u>0$.
Since $\ln^2 \delta/u$ is always positive, $\exp(\ln^2 \delta/u) > 1$.
Since $e^u > u$, eq. (\ref{KM}) follows.
Then
\begin{equation*}
    w(\epsilon) = b v\left(\frac{\epsilon}{b}\right) = \frac{b^2}{\epsilon} \ln\left(\frac{b}{\epsilon}\right),
\end{equation*}
and eq. (\ref{KB}) follows.
With $b = 1.52$, $w(\epsilon) = -2.31 \ln(0.66 \epsilon)/\epsilon$.
\\ \noindent
{\emph{\textcolor{purple}{Lemma 3: Sum over $P(m)$.}}} 
Here we calculate exactly sums of the form
\begin{equation*}
    S(N,\alpha) = \sum_{m=1}^N \alpha^m P(m),
\end{equation*}
where we are specifically interested in $\alpha=2$ and $\alpha = 2^{1 - \epsilon}$.
Inserting $P(m)$, we have
\begin{equation*}
    S(N,\alpha) = \frac{1}{N} N! \sum_{m=1}^N m \frac{1}{(N-m)!} \left(\frac{\alpha}{N}\right)^m.
\end{equation*}
The factor $m$ can be traded for a differentiation with respect to the parameter $\alpha$, giving
\begin{equation*}
S(N,\alpha) = \frac{\alpha}{N} \frac{d}{d\alpha} N! \sum_{m=1}^N \frac{1}{(N-m)!} \left(\frac{\alpha}{N} \right)^m.
\end{equation*}
Increasing the range from $m=1$ to $N$ to $m=0$ to $N$, and swapping $N$ and $N-m$, this can be rewritten
\begin{equation*}
S(N,\alpha) = \frac{\alpha}{N} \frac{d}{d\alpha} 
\left(-1 + \left(\frac{\alpha}{N} \right)^N  N! \sum_{m=0}^N \frac{(N/\alpha)^m}{m!}\right).
\end{equation*}
Since 
\begin{equation*}
N! \sum_{m=0}^N \frac{x^m}{m!} = e^x \Gamma(N+1,x),
\end{equation*}
where $\Gamma(s,x)$ is the upper incomplete gamma function, this becomes
\begin{equation*}
S(N,\alpha) = \frac{\alpha}{N} \frac{d}{d\alpha} 
\left( -1 +  \left( \frac{\alpha e^{1/\alpha}}{N} \right)^N \Gamma(N+1,N/\alpha) \right).
\end{equation*}
Since 
$d/dx \Gamma(N+1,x) = -x^N e^{-x}$, the exact solution is
\begin{equation*}
	S(N,\alpha) = \frac{1}{\alpha} + 
	\left( 1 - \frac{1}{\alpha} \right)
	\left( \frac{\alpha e^{1/\alpha}}{N}\right)^N
	\Gamma(N+1,N/\alpha).
\end{equation*}
In the region we are interested in, $\alpha \leq 2$,
$\Gamma(N+1,N/\alpha)/N!$ rapidly approaches 1 from below. 
Applying Stirling's approximation,
\begin{equation*}
	S(N,\alpha) = \frac{1}{\alpha}+\sqrt{2 \pi N}
	\left( 1 - \frac{1}{\alpha} \right)
	\left( \frac{\alpha}{\exp(1-1/\alpha)}\right)^N.
    \label{SSum}
\end{equation*}
Replacing $\alpha$ with $2^{1-\epsilon}$, and considering the large $N$ limit, the sum has the asymptotic form
\begin{equation*}
	S(N,2^{1\!-\!\epsilon}) \asymp \sqrt{2 \pi N}\left(1\!-\!2^{\epsilon\!-\!1}\right)
	\left(\frac{2^{1\!-\!\epsilon}}{\exp\left(1\!-\!2^{\epsilon\!-\!1}\right)}\right)^N.
\end{equation*}
This gives both of the sums in eq. (\ref{BoundsB}).
\\ \noindent
{\emph{\textcolor{purple}{Lemma 4: Maximum over $\epsilon$.}}} 
We can eliminate $\epsilon$ in eq. (\ref{BoundsD}) by maximising the lower bound with respect to $\epsilon$. 
For large $N$, the dominant terms in the log of the bounds are
\begin{equation}
   \ln \overline{c} 
    \begin{dcases}
        > \! N \! \left(\!(1- \epsilon) \ln 2 + \frac{2^{\epsilon}}{2} - 1 \! \right) +  
        \frac{b^2\ln 2}{\epsilon} \ln \left(\frac{\epsilon}{b}\right)\!, \\
        < \! N \ln \left(\frac{2}{\sqrt{e}}\right).
    \end{dcases}
    \label{LogBoundsWithEpsilon}
\end{equation}
The $\epsilon^*$ that gives the largest lower bound satisfies
\begin{equation*}
N = \frac{2 b^2 \ln \left(\frac{b e }{\epsilon^*}\right)}{{\epsilon^*}^2 \left(2-2^{\epsilon^*}\right) \ln 2}.
\end{equation*}
For example, for $\epsilon^* = 0.01$, $N =$ 278,000.
Since $N$ is a decreasing function of $\epsilon^*$ for $\epsilon^* < 1/2$, 
the maximum must go to 0 as $N$ goes to $\infty$.
\\ \indent
For large $N$, we can replace $2-2^{\epsilon^*}$ with 1.
Doing so allows us to solve for $\epsilon^*$ in terms of $N$:
\begin{equation*}
    \epsilon^* \sim \frac{b \sqrt{W(N e^2)}}{\sqrt{N}},
    \label{EpsilonStar}
\end{equation*}
where $W$ is the principal branch of the Lambert $W$ function.
Inserting this into eq. (\ref{LogBoundsWithEpsilon}), we find
\begin{equation*}
   \ln \overline{c}(N) 
    \begin{dcases}
        > N \ln \left(\frac{2}{\sqrt{e}}\right) 
        \left(1-\frac{5.45}{\sqrt{N}} \frac{(W(e^2 N)-1)}{\sqrt{W(e^2 N)}}
        \right)\!, \\
        <   
        N \ln \left(\frac{2}{\sqrt{e}}\right).
    \end{dcases}
\end{equation*}
where the constant 5.45 is $b/\log_2 (2/\sqrt{e})$. 
Notice that our approach of setting $\epsilon$ to the value that maximizes the lower bound implicitly forces $\epsilon$ to be small for large $N$. 
In retrospect we could have just as well maximized the log of the lower bound in eq. (\ref{BoundsE}) rather than in  in eq. (\ref{BoundsD}).
\\ \\ \noindent
{\sf\textbf{\textcolor{purple}{Discussion}}} \\
The search for the number of attractors in the critical Kauffman network has spanned half a century.
For the model with connectivity one, our work brings this search to a close, because our lower and upper bounds converge to $(2/\sqrt{e})^N$.
If, as many believe \cite{Drossel05a,Drossel05b}, all critical Boolean networks behave in a similar way, 
then our result suggests that the Kauffman model with connectivity two would also exhibit exponential scaling.
\\ \indent
Why has the critical behavior of Kauffman networks proved so elusive, despite the attention of many leading statistical physicists?
Part of the answer lies in the admonition that ``correct scaling
only emerges for very large $N$''~\cite{Socolar03}. 
From eq. (\ref{BoundsB}), for $N = 75$ nodes,
$\overline{c}_{\rm min} = 213$ and $\overline{c}_{\rm max} = 2.12 \times 10^9$.
We know that the true value of $\overline{c}$ lies somewhere in between these, but until $N$ is large, even the logarithms of these numbers are disparate.
From eq. (\ref{BoundsF}), we can write $\overline{c}$ as $x^N$, 
where $x$ ranges between $x_{\rm min}$ and $x_{\rm max} = 1.213$.
This is plotted in Fig. 3.
Not until the network size $N$ is of order 1,000 do the values converge.
\begin{figure}[t!]
\noindent
      \epsfxsize=1 \columnwidth
      \epsfbox{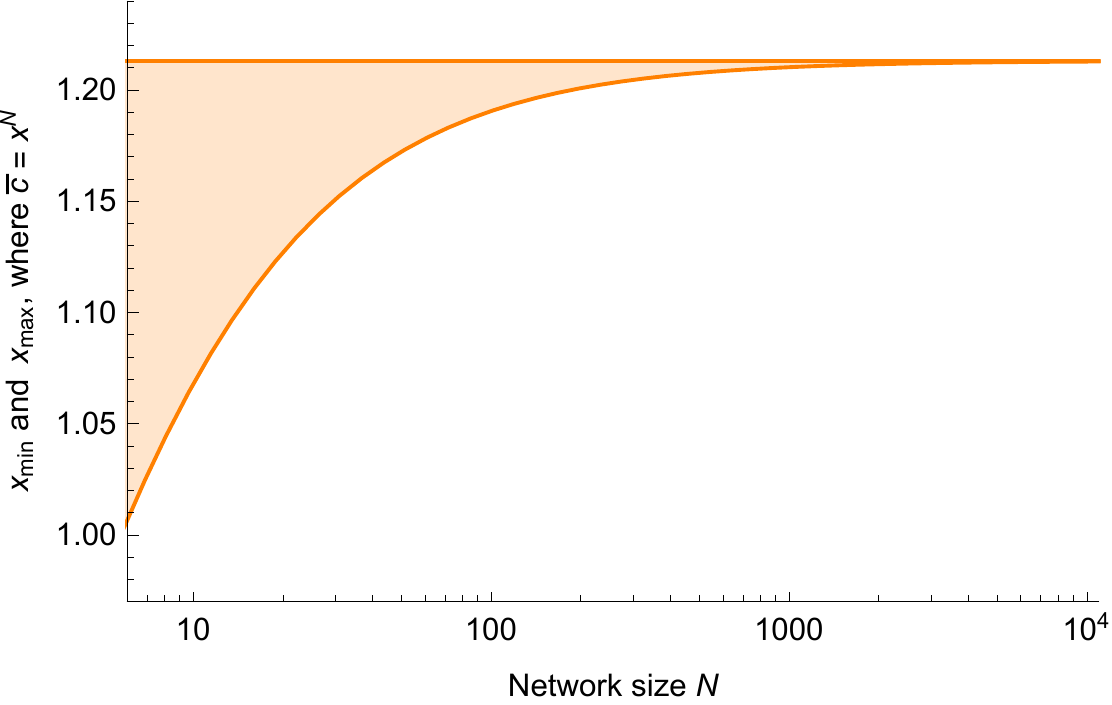}
\begin{small}
\caption{
{\bf Correct scaling requires a large system size.} 
Writing the mean number of attractors as $x^N$, 
here we plot the $x_{\rm min}$ and $x_{\rm max}$, where $x_{\rm max} = 2/\!\sqrt{e} = 1.213$.
The true $x$ is somewhere in the orange area.
The bounds are close only for networks of around $10^3$ nodes.
Notice that the slope of the lines in Fig. 2 are $ \ln x_{\rm min}$ and $\ln x_{\rm max}$.
}
\end{small}
\end{figure}
\\ \indent
That the scaling is exponential in $N$ is surprising as the maximum number of states in attractors is $2^m$ and the mean number of nodes in loops $\ov{m}$ scales as $\sqrt{N}$. Previous work has used Jensen's inequality to obtain the bound $\ov{2^m}>2^{\ov{m}} \sim 2^{\sqrt{N}}$. By computing the expectation exactly, we show that the mean number of attractors scales far more quickly, as $\ov{2^m} \sim (2/\sqrt{e})^N$. This means that the average number of attractors is more than the average model can accomodate, and the expectation is dominated by rare configurations with a number of nodes that is linear in $N$.
\\\indent
While surprising, the result has a familiar basis.
In the large $N$ limit the distribution over $m$ is approximately
$P(m) \approx \frac{m}{N}\exp\left(-m^2/(2N)\right)$.
Taking the expectation $\ov{e^m}$, the familiar completion of the square gives that the average is proportional to $\exp(N)$, rather than $\exp(\sqrt{N})$ in analogy with our exact result.
\\ \indent
Having analytical results for model systems is important for researchers working on applications.
The use of Boolean networks across systems biology~\cite{Li23,Gates21} has inspired computational problems around their control~\cite{Parmer22,Borriello21}, and efficient identification of attractors~\cite{Yuan19}.
Many attractors in the Kauffman model are accessible from only a few states making them difficult to locate through sampling.
As a result ``purely numerical investigation of Kauffman networks will never produce reliable results''~\cite{Kaufman05}.
We hope that the analytical results we present here will be useful to the wider community developing tools and algorithms for Boolean networks.

\begin{footnotesize}

\end{footnotesize}
\end{document}